\documentclass[11pt,a4paper]{article}

\usepackage[verbose=true,letterpaper]{geometry}
\usepackage[font=footnotesize]{caption}

\usepackage[utf8]{inputenc} 
\usepackage[T1]{fontenc}    
\usepackage{amsmath}
\usepackage{epsfig}
\usepackage{graphicx}
\usepackage{wrapfig}
\usepackage{psfrag}
\usepackage{substr}
\usepackage[tight]{minitoc}
\usepackage{subfigure}
\usepackage{longtable}
\usepackage{chngcntr}
\usepackage[all]{xy}
\usepackage{cite}
\usepackage{xr}
\usepackage{etoolbox}

\providecommand{\keywords}[1]
{
	\small	
	\textbf{\textit{Keywords---}} #1
}

\title{Mechanical systems with hyperbolic chaotic attractors based on Froude pendulums}
\date{\normalsize\today}
\author{
	Vyacheslav P.~Kruglov \\
	\small Kotel'nikov Institute of Radioengineering\\ \small and Electronics of RAS\\
	\small Saratov Branch, Russia\\
	\small Udmurt State University\\
	\small Izhevsk, Russia\\
	\small\texttt{kruglovyacheslav@gmail.com} \\
	\and
	Sergey P.~Kuznetsov\\
	\small Kotel'nikov Institute of Radioengineering\\ \small and Electronics of RAS\\
	\small Saratov Branch, Russia\\
	\small Udmurt State University\\
	\small Izhevsk, Russia\\
	\small\texttt{spkuz@yandex.ru} \\
	\and
	Yuliya V.~Sedova \\
	\small Kotel'nikov Institute of Radioengineering\\ \small and Electronics of RAS\\
	\small Saratov Branch, Russia\\
	\small\texttt{sedovayv@yandex.ru}
}

\begin{document}
\maketitle

\begin{abstract}
We discuss two mechanical systems with hyperbolic chaotic attractors of Smale -- Williams type. Both models are based on Froude pendulums. The first system is composed of two coupled Froude pendulums with alternating periodic braking. The second system is 
Froude pendulum with time-delayed feedback and periodic braking. We demonstrate by means of numerical simulations that proposed models have chaotic attractors of Smale -- Williams type. We specify regions of parameter values at which the dynamics corresponds to Smale -- Williams solenoid. We check numerically hyperbolicity of the attractors. 
\end{abstract}


\section{Introduction}

Uniformly hyperbolic attractors~\cite{1} are genuine chaotic and consist only of saddle trajectories. At every point of a uniformly hyperbolic attractor its tangent space can be decomposed into direct sum of two subspaces, stable and unstable. It is important part of definition 
that all possible angles between any vector from stable subspace and any vector from unstable subspace are distanced from zero at every point of the attractor. Uniformly hyperbolic attractors are structurally stable. That means they occupy an open set in the 
parameter space~\cite{1}. 

Smale -- Williams solenoid~\cite{2,3} is well-known example of uniformly hyperbolic chaotic attractor. It appears in the phase space of a descrete dynamical system if a torus-form domain undergoes in one discrete time step an $M$-fold longitudinal stretching 
($M \geq 2$ is integer), strong transverse compression and folding in a loop located inside the initial torus. With each repetition of the transformation, the number of curls increases by factor $M$ and in the limit tends to infinity, resulting in a solenoid with a 
Cantor-like 
transverse structure. Chaotic nature of the dynamics is determined by the fact that the transformation of the angular coordinate in this setup corresponds to an expanding circle map, or the Bernoulli map $\phi _{n+1} = M \phi _n \pmod{ 2 \pi}$. 

Among possible examples of hyperbolic chaos in systems of various nature we outline mechanical models as they are easily perceived and interpreted in a frame of our everyday experience. We propose to consider two mechanical systems~\cite{4,5} with the basic element 
being a self-oscillating Froude pendulum. The first model is composed of two Froude pendulums on a common shaft interacting by friction~\cite{4}. Pendulums are alternately braked by periodic application of external frictional forces. The Smale -- Williams solenoid 
occurs as an attractor of the Poincar\'{e} stroboscopic map as we properly specify the system parameters. The second model is composed of only one Froude pendulum interacting with mechanical time-delay line~\cite{5}. The pendulum undergoes periodic braking. 
The Smale -- Williams solenoid in this model appears in the phase space of infinite dimension of the system with time-delay. 

We start with a quick overview of the Froude pendulum. 

\begin{figure}[!b] 
\includegraphics[width=.49\textwidth,keepaspectratio]{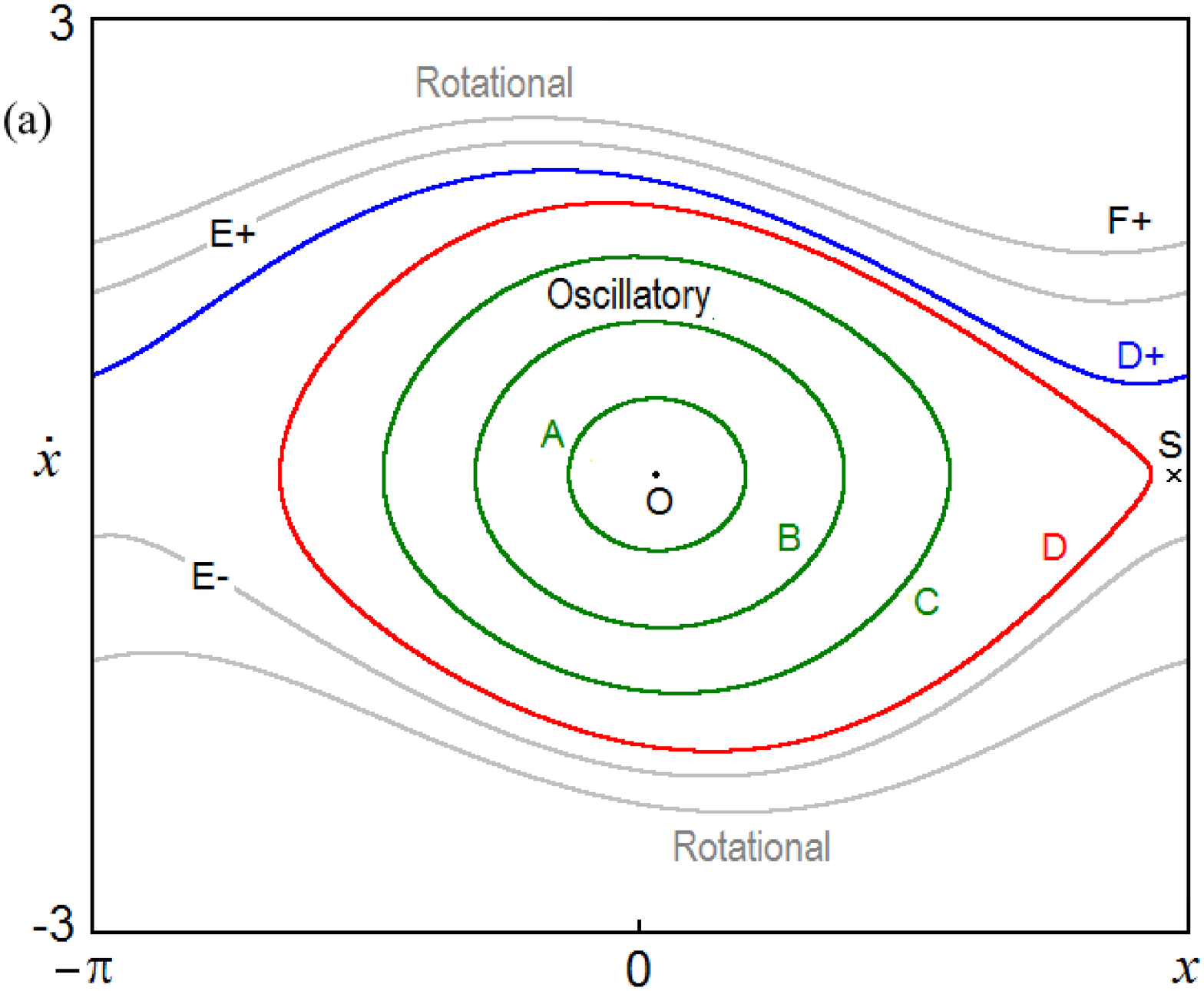}
\includegraphics[width=.49\textwidth,keepaspectratio]{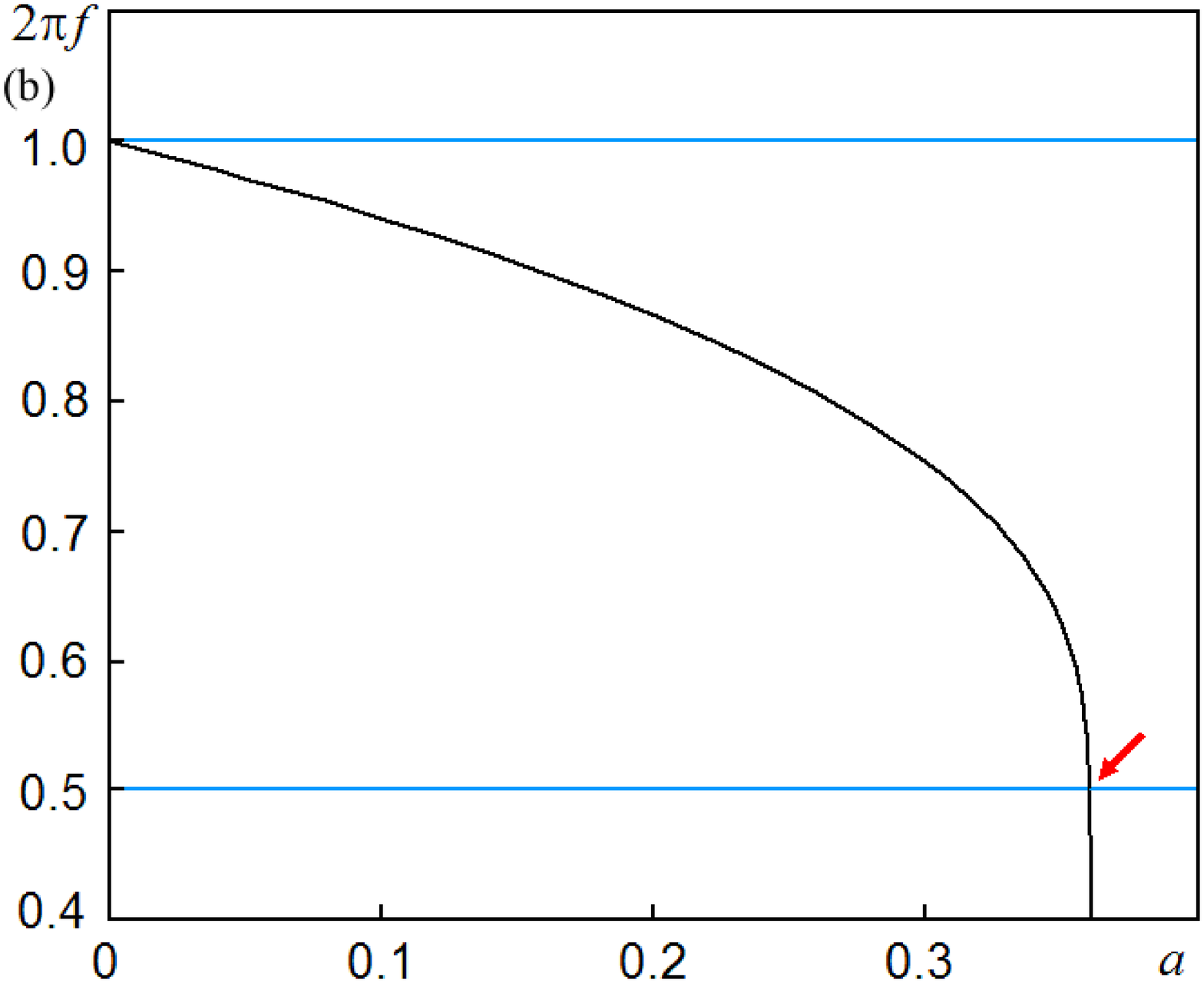}
\caption{(a)~Attracting limit cycles for various parameters corresponding to sustained periodic motions a
single Froude pendulum: oscillatory A, B, C, D respectively, at $a = 0.03$, $0.12$, $0.24$, $0.36$ and rotational D+, E$\pm$, F$\pm$ at $a= 0.36$, $0.48$, $0.6$. (b)~The dependence of the
frequency of the self-oscillating regimes on the value of $a$. Other parameters are $b=0.16$, $\mu=0.087$. On the panel (a) points are marked that are equilibrium states: unstable focus $O$ and saddle $S$. On the panel (b) the arrow indicates the situation where the
frequency of self-oscillations is half the frequency of small oscillations of the pendulum. }
\label{fig01}
\end{figure}

\section{Froude pendulum}

Froude pendulum is a well-known example of mechanical self-oscillator. 
It is a weight on a rod of negligible mass. The rod is attached to a sleeve placed on a shaft rotating at a constant angular velocity. 
In the dimensionless form the governing equation of the Froude pendulum reads
\begin{equation}
\ddot {x} - \left(a - b \dot {x}^2 \right) \dot {x} + \sin x = \mu,
\label{eq4}
\end{equation}
where $x$ is an angle of the pendulum displacement from the vertical line. The friction torque between the shaft and the sleeve gives rise to self-oscillatory motions of the Froude pendulum at certain range of parameters.

If $a>0$, then self-oscillations arise in the system. On the phase plane Fig.~\ref{fig01}~(a) the self-oscillatory modes are represented by limit cycles around the equilibrium state $O$ at 
$\left(x, \dot{x} \right) = \left( \arcsin \mu, 0 \right)$. For $a$ near zero and small $\mu$, a frequency of the self-oscillations is close to the natural frequency of the pendulum $f=(2\pi)^{-1}$. With growth of $a$, the limit cycles increase in size, and the frequency $f$
decreases. This is due to the fact that the pendulum approaches the saddle point $S$, $\left(x, \dot{x} \right) = \left(\pi - \arcsin \mu, 0 \right)$, corresponding to position pointing upwards, where the motion  slows down. Further growth of 
the parameter $a$ leads to a change of the oscillatory movements of the pendulum to periodic rotational motions, which correspond to the limit cycles going around the phase cylinder. The dependence of the frequency of the self-oscillating regimes on the value of $a$ is
shown on Fig.~\ref{fig01}~(b).

For further consideration, we select parameters to obtain self-oscillations with the frequency exactly equal to half of the frequency of small self-oscillations at $a$ close to zero. At $\mu=0.087$ and $b=0.16$ this is the case if we set $a=0.36$. Then the oscillatory process 
has an essential second harmonic of the fundamental frequency (this is due to the lack of symmetry of the equation~\eqref{eq4} with respect to the substitution $x \leftrightarrow -x$). If the generated signal acts on a linear oscillator of natural frequency $\omega=1$, one
can observe its resonant excitation under the second harmonic of the self-oscillating system.

\section{System of two Froude pendulums with alternating periodic \mbox{braking}}

Let us consider two identical Froude pendulums placed on a common shaft and weakly connected with each other by viscous friction, so that the torque of the frictional force is proportional to the relative angular velocity. Let the motion of pendulums
be decelerated alternately by attaching a brake shoe providing suppression of the self-oscillations due to the incorporated sufficiently strong viscous friction. Denoting the angular coordinate of the first and the second pendulum as $x$ and $y$, and the angular 
velocities as $u$ and $v$, we write down the equations 

\begin{equation}
\begin{array}{l}
 \dot {x} = u, \\
 \dot {u} = \left[ a - d \left(t\right) - b u^2 \right] u - \sin x + \mu + \varepsilon \left(v - u \right), \\
 \dot {y} = v, \\
 \dot {v} = \left[ a - d \left(t + T / 2 \right) - b v^2 \right] v - \sin y + \mu + \varepsilon \left(u - v\right), \\ \\
 d(t) = \left\{ {{\begin{array}{*{20}c}
 {0,\,\,t < T_0 ,\,\,\,\,\,\,\,\,\,\,\,\,\,\,\,\,\,} \hfill \\
 {D,\,\,T_0 < t < T / 2,} \hfill \\
 {0,\,\,T / 2 < t < T.\,\,} \hfill \\
\end{array} }} \right.\,\,\,\,\,\,d(t + T) = d(t). \\
 \end{array}
 \label{eq6}
\end{equation}
Parameters are assigned as follows:
\begin{equation}
\begin{array}{c}
a = 0.36,\,\,b = 0.16,\,\,\mu = 0.087,\,\,\varepsilon = 0.0003,\,\,D = 0.8,\,\,T = 250,\,\,T_0 = T / 4.
\end{array}
\label{eq7}
\end{equation}

To explain the operation of the system~\eqref{eq6} we start with the situation when one pendulum is self-oscillating, and the second is suppressed to small oscillations by brake. Due to the fact that the parameters are chosen in accordance with the reasoning of the 
previous section, the basic frequency of the self-oscillatory mode is half of that of the second pendulum. Therefore, when the brake shoe is removed from the second pendulum, it will begin to swing in a resonant manner due to the action of 
the second harmonic from the first pendulum, and the phase of the oscillations that arise will correspond to the doubled phase of the oscillations of the first pendulum. As a result, when the second pendulum approaches the sustained self-oscillatory state,
its phase appears to be doubled in comparison with the initial phase of the first pendulum. Further, the first pendulum undergoes braking, and at the end of this stage, its oscillations will be stimulated in turn by the action of the second harmonic from the 
second pendulum, and so on.

Since the system~\eqref{eq6} is non-autonomous, one can go on to discrete time dynamics by constructing the Poincar\'{e} stroboscopic map. In our case, taking into account a symmetry of the system in respect to substitution
$x \leftrightarrow y,\,\,u \leftrightarrow v,\,\,t \leftrightarrow t + T / 2$, it is appropriate to use the mapping in half a period of modulation, determining the state vector at the instants of time as
\begin{equation}
\label{eq8}
{\rm {\bf X}}_n = \left\{ {{\begin{array}{*{20}c}
 {(x(t_n ),\,u(t_n ),\,y(t_n ),\,v(t_n )),\,\,\mbox{if}\,\,n\,\,\mbox{is
odd},\,} \hfill \\
 {(y(t_n ),\,v(t_n ),\,x(t_n ),\,u(t_n )),\,\,\mbox{if}\,\,n\,\,\mbox{is
even.}} \hfill \\
\end{array} }} \right.
\end{equation}
The Poincar\'{e} map for the vector ${\rm {\bf X}}_n = (x_1 ,x_2 ,x_3 ,x_4 )_n$:
\begin{equation}
\label{eq9}
{\rm {\bf X}}_{n + 1} = {\rm {\bf F}}_{T / 2} ({\rm {\bf X}}_n ).
\end{equation}

\begin{figure}[!h]
\includegraphics[width=.52\textwidth,keepaspectratio]{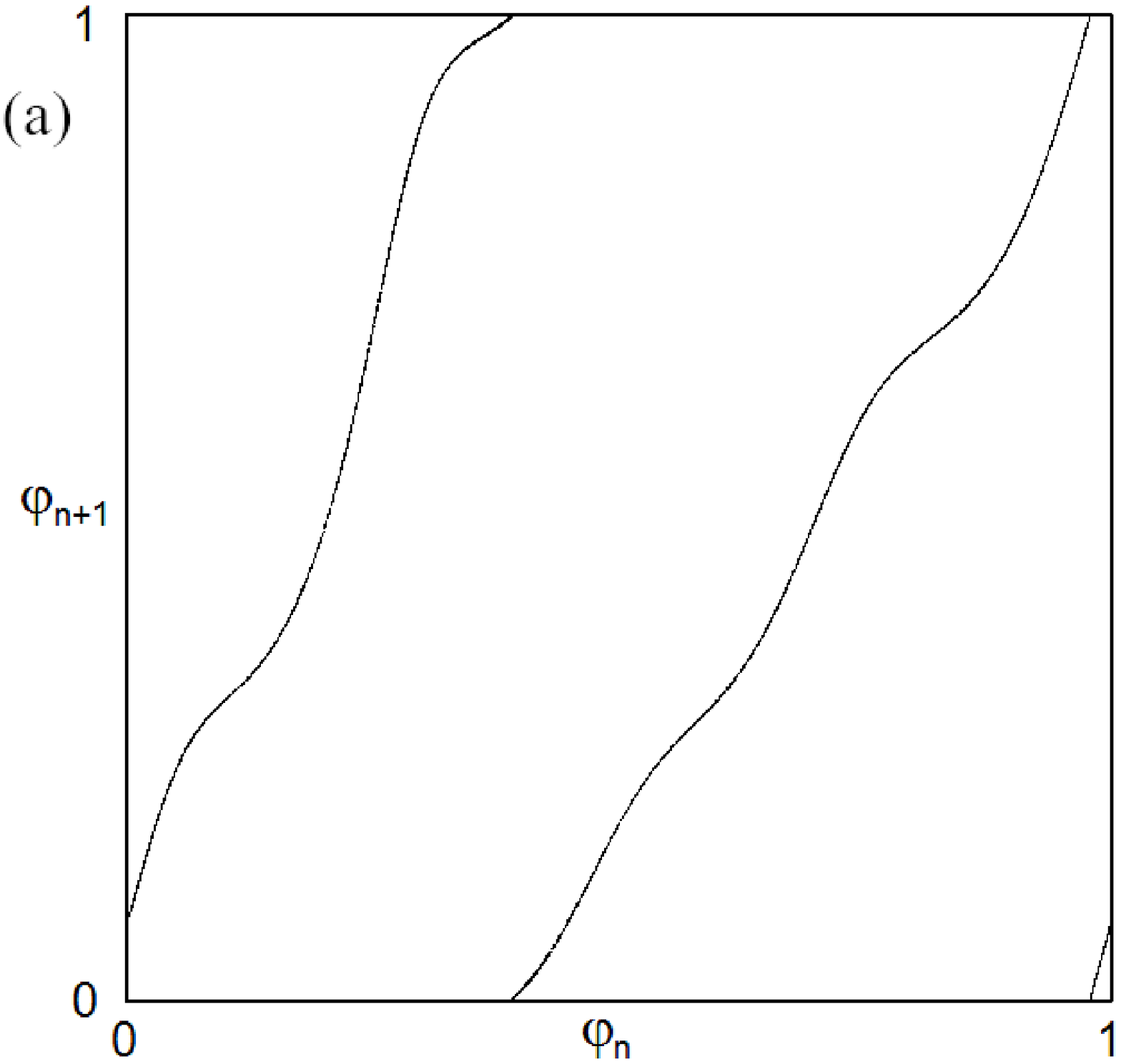}
\includegraphics[width=.48\textwidth,keepaspectratio]{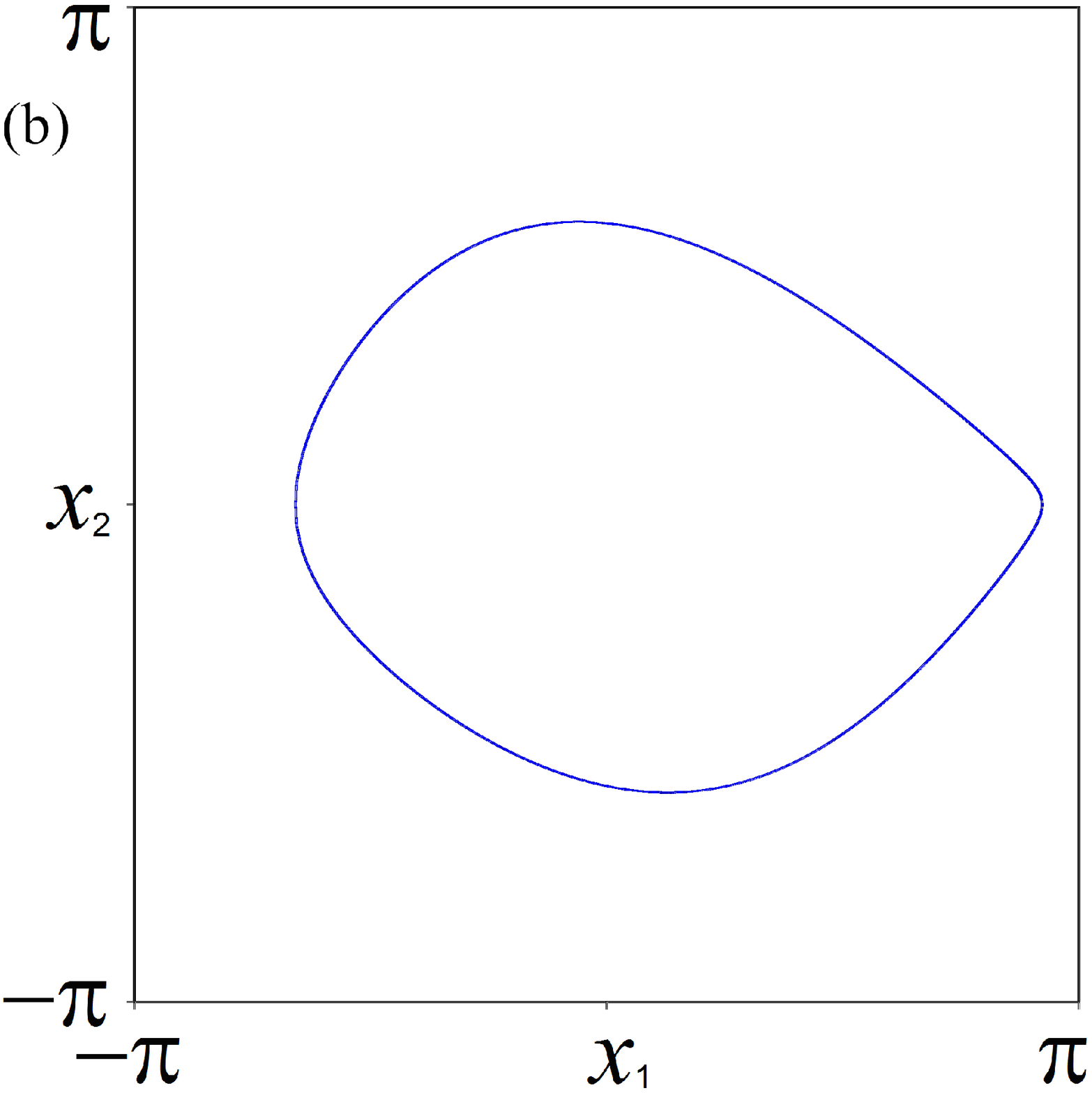}
\caption{(a)~A diagram illustrating transformation of the phases of pendulums
in successive stages of activity every half a period of modulation. (b)~Attractor of the Poincar\'{e} stroboscopic map, which is a
Smale -- Williams solenoid in projection onto the plane of two of variables. }
\label{fig02}
\end{figure}

Since each new stage of the excitation transfer to one or another pendulum is accompanied by a doubling of the phase of oscillations, this corresponds to the Bernoulli map for the phase. If a volume contraction takes place along the
remaining directions in the state space of the system, this will correspond to occurrence of the Smale -- Williams solenoid as attractor of the
Poincar\'{e} map~\eqref{eq9}.

At the stage of the high-amplitude oscillations their waveform differs significantly from the sinusoidal. In this case, evaluation of the phase through the ratio of the variable and its derivative as arctangent is not so satisfactory. We define the phase using a value of the time shift of the waveform with respect to a given reference point, normalized to the characteristic period of the self-oscillatory mode. Let $t$ be a fixed time instant 
at the activity stage of one of the pendulums, and $t_{1}$, $t_{2}$ are the preceding moments of the sign change of the angular velocity from plus to minus, and $t_2 > t_1 $. Then we can define the phase as a variable belonging to the interval [0,~1]
by the relation $\varphi = (t - t_2 )(t_2 - t_1 )^{ - 1}$.

Fig.~\ref{fig02}~(a) shows a diagram for the phases determined at the end parts of successive stages of excitation of the first and second pendulums, obtained in numerical calculations for a sufficiently large number of the modulation periods. As can be seen, the 
mapping for the phase 
in the topological sense looks equivalent to the Bernoulli map $\varphi _{n + 1} = 2\varphi _n + \mbox{const}\,\,(\bmod \,1)$. Indeed, one complete round for the pre-image $\varphi _n $ (i.e., a unit shift) corresponds to a double round for the image 
$\varphi _{n + 1} $.

Fig.~\ref{fig02}~(b) shows attractor of the Poincar\'{e} stroboscopic map. Although visually the object looks like a closed curve, in fact it has a fine transverse structure, visualization of which requires high-accuracy calculations, and evolution in discrete time 
corresponds to jumps of the representing point around the loop accordingly to iterations of the Bernoulli map. 

At the parameters assigned according to~\eqref{eq7}, the Lyapunov exponents for the Poincar\'{e} map attractor are
\[
\Lambda _1 = 0.65\pm 0.007,\, \Lambda _2 = - 5.89\pm 0.02,\, 
\Lambda _3 = - 9.46\pm 0.02,\, \Lambda _4 = - 20.73\pm 0.02.
\]
The presence of a positive exponent $\Lambda _{1}$ indicates chaotic nature of the dynamics. Its value is close $\ln 2 = 0.693 \ldots $, which agrees with the approximate description of the evolution of the phase variable $\varphi$ by the Bernoulli map. The action 
of the Poincar\'{e} map \textbf{F} in the four-dimensional space is accompanied by stretching in the direction corresponding to the phase $\varphi$ and contracting along the remaining three directions. This corresponds to the Smale -- Williams construction,
namely, in the four-dimensional space. 

\begin{figure}[!h]
\includegraphics[width=.55\textwidth,keepaspectratio]{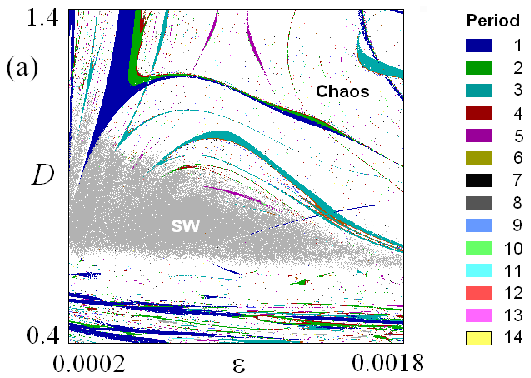}
\includegraphics[width=.44\textwidth,keepaspectratio]{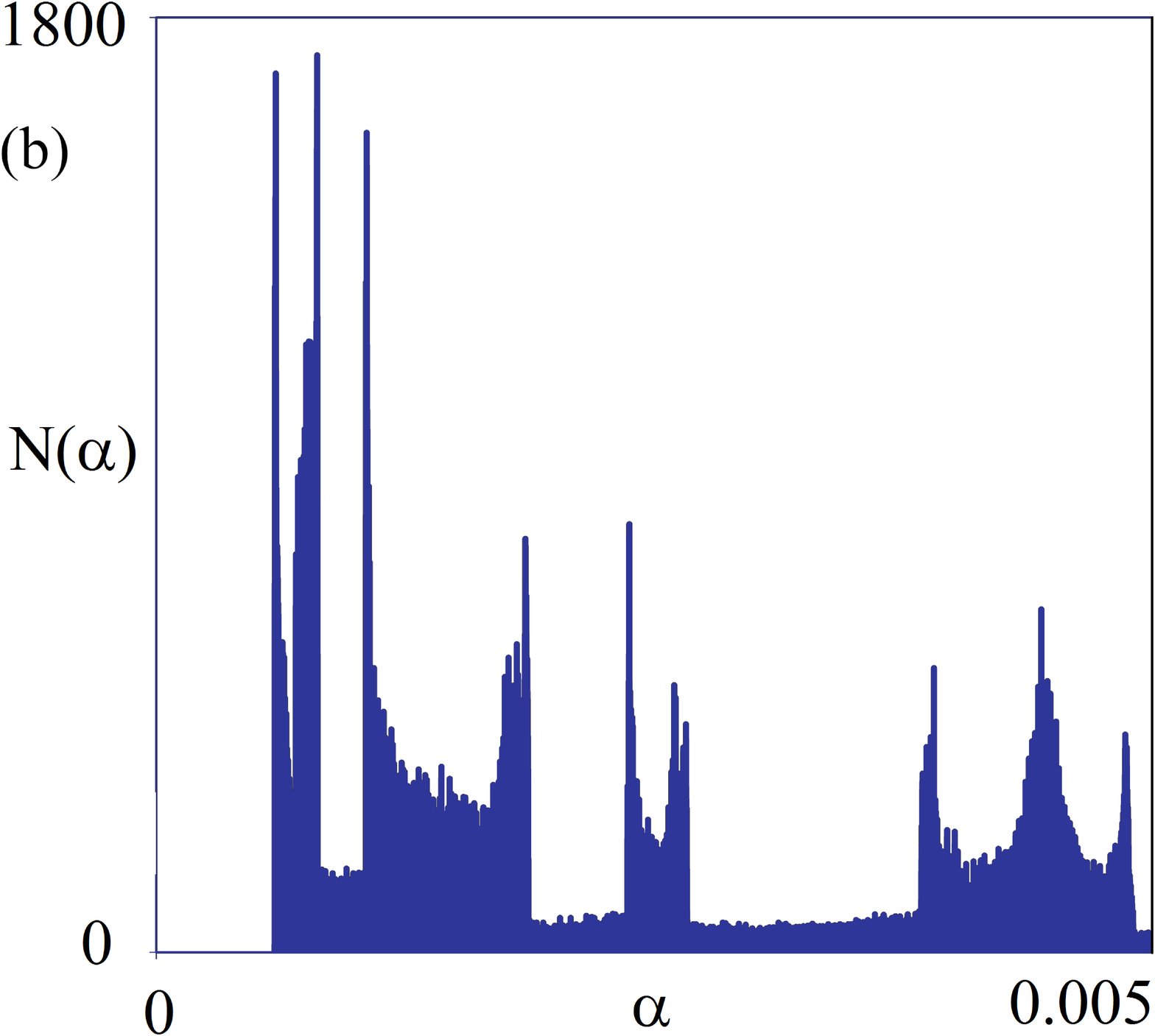}
\caption{(a) Chart of regimes of the system~\eqref{eq6} on the plane $(\varepsilon, D)$, where areas of chaos and of periodic motions are shown. The
region of hyperbolic chaos associated with the Smale -- Williams attractor is shown in gray and marked as SW, other chaotic
regimes are shown in white. The legend for periodic regimes is shown on the right. The periods 
indicated in colors are measured in units of modulation period. Fixed parameters are $a = 0.36$, $b = 0.16$, $\mu = 0.087$, $T = 250$, $T_0 = T / 4$.
(b) Histogram of the angles of intersection of stable and unstable subspaces for the hyperbolic attractor of the Poincar\'{e} map of the system~\eqref{eq6}. $a = 0.36$, $b = 0.16$, $\mu = 0.087$, $\varepsilon = 0.0003$, $D = 0.8$, $T = 250$, $T_0 = T / 4$.}
\label{fig03}
\end{figure}

Fig.~\ref{fig03}~(a) shows a chart of regimes on the parameter plane $(\varepsilon, D)$, i.e. of the coupling parameter versus the dissipation parameter introduced by the brake pad during the braking stages. The remaining parameters correspond to the situation when 
the period of relaxation self-oscillations at the activity stage is exactly twice the period of small oscillations. The structure of the regions is determined by the excitation transfer from one stage of activity to the next one in the course
of the system operation. In the central part of the chart one can see a broad area SW, where hyperbolic chaos takes place. To identify Smale -- Williams solenoid the topological equivalence of the map for the phase to the Bernoulli map was tested visually or 
automatically, using a specially developed algorithm. Exit this area down corresponds to the fact that dissipation at the stages of breaking decreases and becomes too small to provide a sufficient 
degree of damping of the natural oscillations of the pendulum after the previous activity stage, which makes a competing contribution to the stimulation of the oscillatory process at the new activity stage, so that the phase doubling transfer mechanism is violated. The 
exit through the upper boundary of the SW region corresponds to the fact that at the curve representing the graph for the phases, a bend is formed on one of the branches, and then a maximum and a minimum appear, so that the monotony property is lost. This means that 
when a stretched double loop is inserted into the original toroidal region in the solenoid construction procedure, there appears a local fold on the loop, which leads to disruption of the proper solenoid structure. When the parameters at the upper edge of the region SW 
are varied, periodic motions become possible; on the chart one can see there a set of periodicity tongues. Visually, they look similar to the classical synchronization Arnold tongues, but the principal difference from the classical picture is that between them we have
chaotic dynamics rather than the quasiperiodicity.

Fig.~\ref{fig03}~(b) shows the histogram of the angles of intersection of stable and unstable subspaces for a trajectory on the attractor of the Poincar\'{e} map of the system with parameters assigned according to~\eqref{eq7}. The fact that the
distribution is separated from zero, confirms the hyperbolic nature of the attractor. Similar results are obtained for parameters of the system in a certain range, which corresponds to the structural stability inherent to the hyperbolic attractor.
The technique of the performed numerical test was developed in~\cite{6}. 

\section{Froude pendulum with delayed feedback and periodic \\ \mbox{braking}}

Let us consider a Froude pendulum placed on a rotating shaft and contacted with mechanical time-delay transmission line (a spring with a free end). Let the motion of pendulum
be decelerated periodically by attaching a brake shoe. Denoting the angular coordinate of the  pendulum as $x$, and the angular velocity as $u$, we write down the equations
\begin{equation}
\label{eq10}
\begin{array}{l}
 \dot {x} = u, \\
 \dot {u} = [a - d(t) - b u^2]u - \sin x + \mu + \varepsilon \left( u \left( t - \tau \right) - u \left( t \right) \right), \\ \\
 d(t) = \left\{ {{\begin{array}{*{20}c}
 {0,\,\,t < T_0 ,\,\,\,\,\,\,\,\,\,\,\,\,\,\,\,\,\,} \hfill \\
 {D,\,\,T_0 < t < T / 2,} \hfill \\
 {0,\,\,T / 2 < t < T.\,\,} \hfill \\
\end{array} }} \right.\,\,\,\,\,\,d(t + T) = d(t). \\
 \end{array}
\end{equation}

The system~\eqref{eq10} is non-autonomous, with periodic parameter modulation, so we can use the description of the dynamics in terms of discrete time by means of the stroboscopic Poincar\'{e} map
\begin{equation}
\label{eq11}
\rm{\bf{X}}_{n} = {\rm {\bf F}}_{T} \left(\rm{\bf{X}}_{n-1} \right).
\end{equation}
Here vectors $\rm{\bf{X}}_{n}$ denote the sets of quantities $x(t_n)$, $\dot{x} \left( t_n \right)$ together with functions $\dot{x} \left( t - \tau \right)$, $t \in \left[ t_n - \tau,\,t_n \right) $ determining the state of the system at the time instants 
$t_n = n T$, and should be interpreted as elements of infinite-dimensional state space.

Parameters are assigned as follows:
\begin{equation}
\begin{array}{c}
a = 0.36,\,b = 0.16,\,\mu = 0.087,\,\varepsilon = 0.0003,\\ D = 0.8,\,T = 250,\,T_0 = T / 4,\,\tau = T / 2.
\end{array}
\label{eq12}
\end{equation}

Let us start with a situation when the braking is not applied, and the pendulum performs relaxation self-oscillations, in which, due to the selection of the parameters, the frequency is half of that for the small oscillations of the pendulum. The signal
generated at this stage is being sending to the time-delay feedback transmission line. Further, the pendulum oscillations are suppressed by application of the brake pad. When braking stops, a new stage of the buildup of the
oscillations begins, starting from the practically unexcited state of the pendulum. As we properly select the delay time, the oscillatory process will be stimulated by resonant action of the second harmonic of the signal received through the 
feedback circuit and emitted just on the previous stage of intense oscillations. Therefore, the phase of the developing oscillations corresponds to the doubled phase of the main component of oscillations at the prior activity stage. As a result, when the newly 
arisen oscillations of the pendulum approach the sustained relaxation self-oscillations, their phase will be doubled compared to the phase at the preceding activity stage. Further, the process is repeated again and again.
A full cycle corresponding to the modulation period $T$ is accompanied by multiplying the initial phase of the oscillatory process by factor of $2$, i.e. for it, a doubly expanding circle map takes place. Due to compression in
remaining directions in the state space of the map~\eqref{eq11}, the Smale -- Williams attractor arises.

Numerical simulations were undertaken. It was demonstrated that dynamics of phases indeed is described by 
a map topologically equivalent to the Bernoulli mapping. The calculation of the largest three Lyapunov exponents of the Poincar\'{e} map for the attractor at the given parameters~\eqref{eq12} yields
\[
\Lambda _1 = 0.65,\, \Lambda _2 = - 6.325,\, \Lambda _3 = - 7.061.
\]
The positive value of the first exponent indicates chaotic nature of the dynamics, and, as we can see, it is quite close to the value $\ln 2 = 0.693 \ldots$ (Lyapunov exponent of the Bernoulli map). Other Lyapunov exponents are negative. This corresponds to the 
Smale -- Williams solenoid embedded in the infinite-dimensional space. 

\begin{figure}[!h]
\includegraphics[width=.57\textwidth,keepaspectratio]{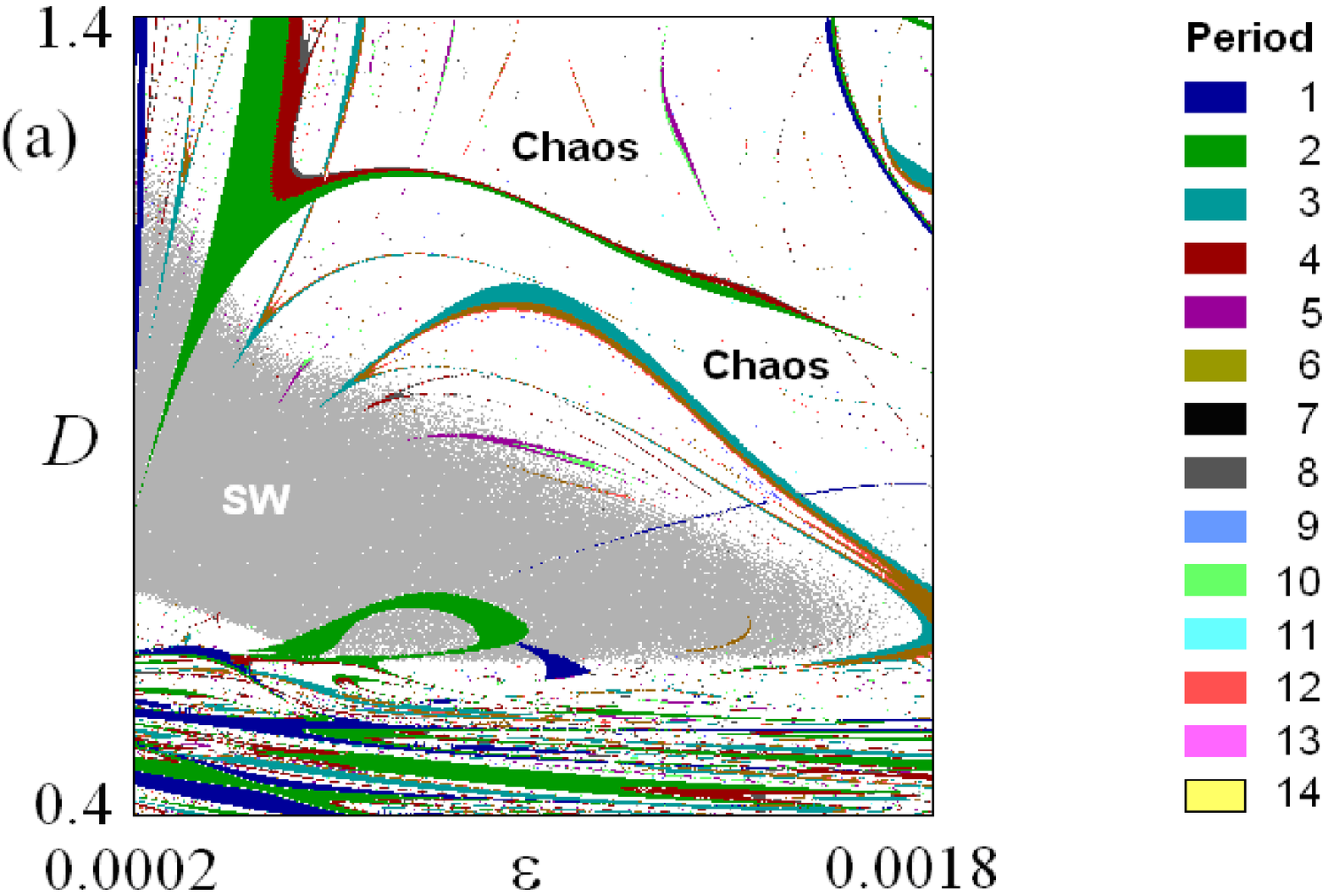}
\includegraphics[width=.43\textwidth,keepaspectratio]{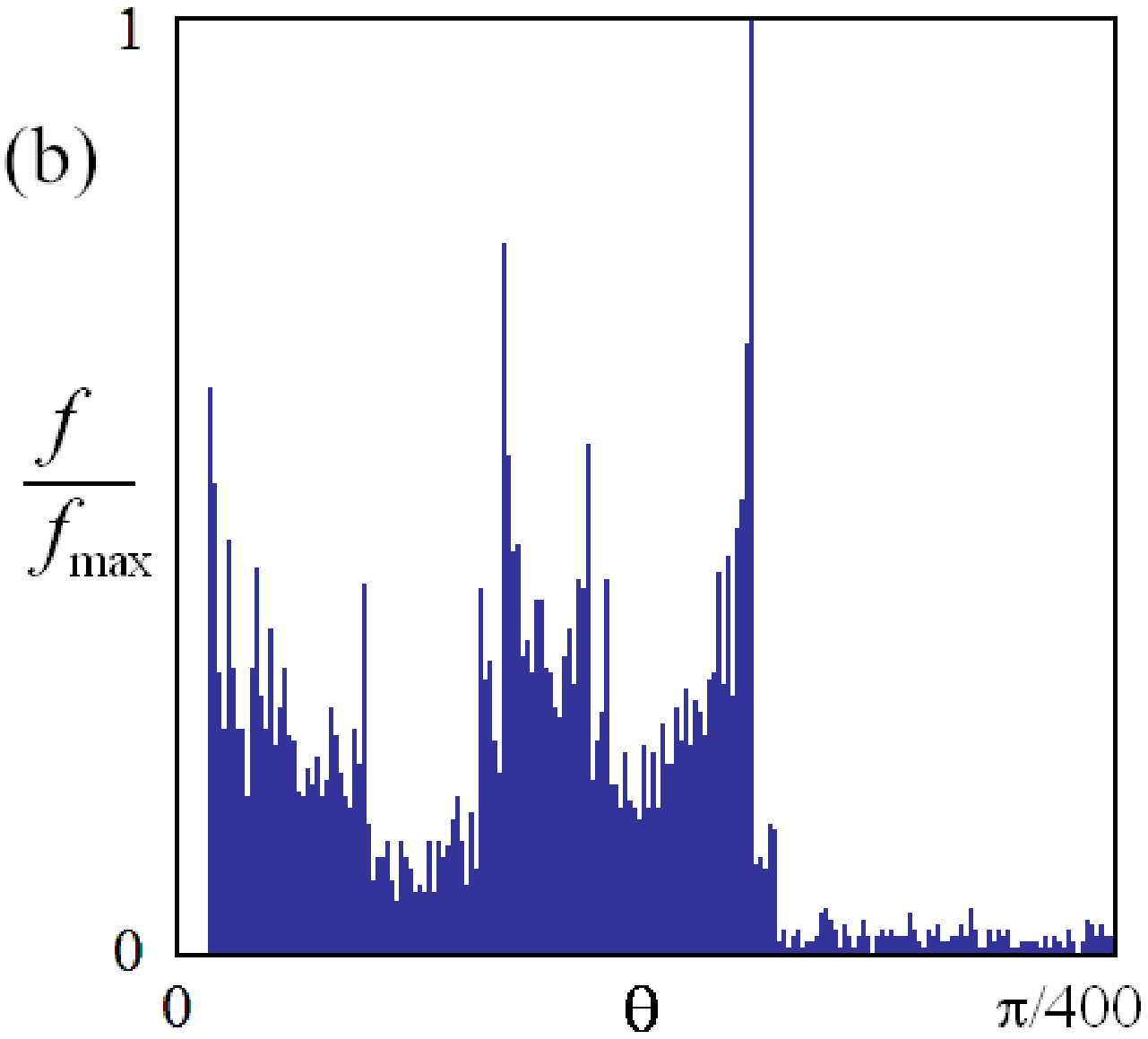}
\caption{(a) Chart of regimes of the system~\eqref{eq10} on the plane $(\varepsilon, D)$, where areas of chaos and of periodic motions are shown. The
region of hyperbolic chaos associated with the Smale -- Williams attractor is shown in gray and marked as SW, other chaotic
regimes are shown in white. The legend for periodic regimes is shown on the right. 
The periods indicated in colors are measured in units of the modulation period.
Fixed parameters are $a = 0.36$, $b = 0.16$, $\mu = 0.087$, $T = 250$, $T_0 = T / 4$, $\tau = T / 2$.
(b) Histogram of the angles of intersection of stable and unstable subspaces for the hyperbolic attractor of the Poincar\'{e} map of the system~\eqref{eq10}. $a = 0.36$, $b = 0.16$, $\mu = 0.087$, $\varepsilon = 0.0003$, $D = 0.8$, $T = 250$, $T_0 = T / 4$, 
$\tau = T / 2$.}
\label{fig04}
\end{figure}

Fig.~\ref{fig04}~(a) shows a chart of regimes on the parameter plane $(\varepsilon, D)$, that is the time-delay feedback intensity parameter versus the dissipation parameter introduced by the brake pad during the braking stages. One can see 
a broad domain SW of hyperbolic chaos. Observe remarkable similarity of the chart to that in Fig.~\ref{fig03}; the difference is distinct coloring of the periodicity areas, as in the system of coupled pendulums the phase transfer happens twice on the modulation period, 
while in the time-delay system it takes place one time on a modulation period.
Fig.~\ref{fig04}~(b) shows the histogram of the angles of intersection of stable and unstable subspaces for a trajectory on the attractor of the Poincar\'{e} map of the system~\eqref{eq10} with parameters assigned according
to~\eqref{eq12}. The distribution is distanced from zero, that confirms the hyperbolicity of the attractor. The special method of numerical test of hyperbolicity for time-delayed systems was developed in~\cite{7}. 

\section{Conclusion}

We discussed two mechanical models based on Froude pendulum with Smale -- Williams hyperbolic attractors. Both models were simulated numerically. Areas of hyperbolic dynamics were identified in the parameter space checking the topological nature of the map for the 
angular variable. The hyperbolicity of the chaotic attractors was also confirmed using a criterion based on the analysis of the intersection angles of stable and unstable invariant subspaces of small perturbation vectors and checking the absence of tangencies between 
these subspaces.

\section*{Funding}

S. P. Kuznetsov and V. P. Kruglov acknowledge support of the grant of Russian Science Foundation No. 15-12-20035 (Sections 1-3). Yu.V. Sedova acknowledges support of the grant of Russian Science Foundation No. 17-12-01008 (Section 4).

\bibliographystyle{unsrt}

\begin{thebibliography}{99} \itemsep=4pt

\vspace{-1mm}

\bibitem{1} Shilnikov, L., Mathematical problems of nonlinear dynamics: a tutorial, \textit{International Journal of Bifurcation and Chaos}, 1997, vol.\,7, no.\,09, pp.\,1953--2001.
\bibitem{2} Smale, S., Differentiable Dynamical Systems, \textit{Bull. Amer. Math. Soc.}, 1967, vol.\,73, no.\,6, pp.\,747--817.
\bibitem{3} Williams, R., Expanding Attractors, \textit{Inst. Hautes Etudes Sci. Publ. Math.}, 1974, no.\,43, pp.\,169--203.
\bibitem{4} Kuznetsov, S.P. and Kruglov, V.P., Hyperbolic chaos in a system of two Froude pendulums with alternating periodic braking, \textit{Communications in Nonlinear Science and Numerical Simulation}, 2019, vol.\,67, pp.\,152--161.
\bibitem{5} Kuznetsov, S.P. and Sedova, Yu.V., Robust hyperbolic chaos in Froude pendulum with delayed feedback and periodic braking, \textit{IJBC}, 2019, DOI:10.1142/S0218127419300350, accepted. 
\bibitem{6} Kuptsov, P.V., Fast numerical test of hyperbolic chaos, \textit{Phys. Rev. E}, 2012 vol.\,85, 015203.
\bibitem{7} Kuptsov, P.V. and Kuznetsov, S.P., Numerical test for hyperbolicity of chaotic dynamics in time-delay systems, \textit{Phys. Rev. E}, 2016 vol.\,94, 015203.

\end{thebibliography}

\end{document}